\pdfoutput=1

\documentclass[11pt, table, a4paper]{article}

\usepackage[]{ACL2023}

\usepackage[ruled,linesnumbered]{algorithm2e}

\usepackage{times}
\usepackage{latexsym}
\usepackage{graphicx}
\usepackage{soul}
\usepackage{booktabs}
\usepackage{color}
\usepackage{float}

\usepackage[noend]{algpseudocode}

\usepackage{longtable}
\usepackage{rotating}
\usepackage{xcolor}

\usepackage{multirow}
\usepackage{colortbl}
\usepackage{hhline}
\usepackage{amssymb}

\usepackage[T1]{fontenc}

\usepackage[utf8]{inputenc}

\usepackage{microtype}

\usepackage{inconsolata}

\newcommand{\tool}{\textsc{Steps}}

\title{Interactive Text-to-SQL Generation via Editable Step-by-Step Explanations}

\author{
Yuan Tian$^1$,
{ Zheng Zhang$^2$},
{ Zheng Ning$^2$},
\\
\textbf{Toby Jia-Jun Li$^2$},
\textbf{ Jonathan K. Kummerfeld$^3$,  and}
\textbf{Tianyi Zhang$^1$} \\
Purdue University$^1$,
{ University of Notre Dame$^2$,}
{ The University of Sydney$^3$} \\
\texttt{tian211@purdue.edu}, 
\texttt{ zzhang37@nd.edu}, 
\texttt{ zning@nd.edu}, \\
\texttt{ toby.j.li@nd.edu}, 
\texttt{ jonathan.kummerfeld@sydney.edu.au}, 
\texttt{ tianyi@purdue.edu}
}

\begin{document}
\maketitle

\begin{abstract}


Relational databases play an important role in business, science, and more.
However, many users cannot fully unleash the analytical power of relational databases, because they are not familiar with database languages such as SQL.
Many techniques have been proposed to automatically generate SQL from natural language, but they suffer from two issues: (1) they still make many mistakes, particularly for complex queries, and (2) they do not provide a flexible way for non-expert users to validate and refine incorrect queries.
To address these issues, we introduce a new interaction mechanism that allows users to directly edit a step-by-step explanation of a query to fix errors. 
Our experiments on multiple datasets, as well as a user study with 24 participants, demonstrate that our approach can achieve better performance than multiple SOTA approaches. 
Our code and datasets are available at \url{https://github.com/magic-YuanTian/STEPS}.



\end{abstract}

\section{Introduction}

Natural language interfaces significantly lower the barrier to accessing databases and performing data analytics tasks for users who are not familiar with database query languages. Many approaches have been proposed for generating SQL queries from natural language  
\citep{popescu-etal-2004-modern, giordani-moschitti-2012-translating, smbop, picard, ship}.
Using recent large language models, systems have reached 86.6\% execution accuracy \cite{gao2023texttosql} on the Spider benchmark \citep{spider}.

\begin{figure}
    \centering
    \includegraphics[width=0.9\linewidth]{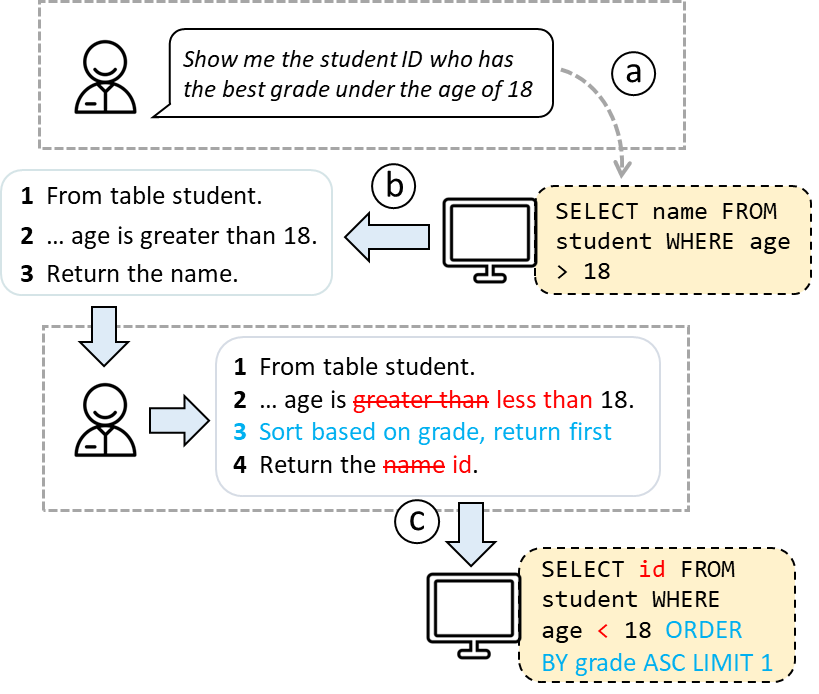}
    \caption{Refining a SQL query by directly editing a step-by-step explanation.}
    \label{fig:idea}
\end{figure}

However, the rate of improvement has slowed, with a gain of only 10\%
since mid-2021. This is partly due to the inherent ambiguity of natural language and the complex structure of SQL queries (e.g., nested or joined queries). 
Thus, it is challenging to generate a fully correct query in one step, especially for complex tasks \citep{misp}. 


There has been growing interest in developing ``human-in-the-loop'' approaches that elicit user feedback to guide SQL generation.
However, most approaches only support feedback in 
constrained forms, e.g., answering multiple-choice questions~\citep[MISP, PIIA, DialSQL][]{misp, piia, dialsql}, changing SQL elements in a drop-down menu~\citep[DIY,][]{diy}, etc. Such constrained feedback is not sufficient to fix many complex errors in real-world SQL tasks.
One exception is NL-EDIT~\citep{nledit}, 
which allows users to provide feedback as new utterances. However, since the feedback is open-ended, interpreting it can be just as hard as processing the original request.



In this paper, we seek to strike a balance between constrained feedback and open-ended feedback by proposing a new interaction mechanism: editable step-by-step explanations. Fig.~\ref{fig:idea} illustrates our idea.
This mechanism consists of three core components: (a) a text-to-SQL model, (b) an explanation generation method, and (c) a SQL correction model.
Our key insight is that using a step-by-step explanation as the basis to suggest fixes allows users to precisely specify where the error is and how to fix it via direct edits. This not only saves users' time 
but also makes it easier for the model to locate the error and apply fixes.



Based on this idea, we implemented an interactive SQL generation and refinement system called {\tool}. 
{\tool} adopts a rule-based method to generate step-by-step explanations and uses a hybrid rule/neural method to convert a user-corrected explanation back to a SQL query. 

An evaluation with a simulated user on Spider~\citep{spider} shows that {\tool} can achieve 97.9\% exact set match accuracy, outperforming prior interactive text-to-SQL systems---MISP, DIY, and NL-EDIT---by 33.5\%, 33.2\%, and 31.3\% respectively.
We further evaluate {\tool} on other datasets, including Spider-DK~\citep{spider-dk}, Spider-Syn~\cite{spider-syn}, and WikiSQL~\citep{wikisql}. {\tool} consistently achieves at least 96\% exact set match accuracy and execution accuracy across all datasets.

Finally, we conducted a within-subjects user study with 24 real users.
We found that within the same amount of time, {\tool} helped users complete almost 2X and 4X more tasks correctly than DIY and MISP respectively,\footnote{We worked with the authors of NL-EDIT to include their system in the user study, but were unable to get it working due to missing code and other runtime errors. We use the accuracy reported in the NL-EDIT paper for comparisons.} with significantly higher self-reported confidence and lower mental load.

This work makes the following contributions: (1) we propose a new interaction mechanism for the text-to-SQL task; (2) we develop an interactive text-to-SQL system based on the new interaction mechanism and a new training method for SQL correction; (3) we conduct a comprehensive evaluation with both simulated and real users and demonstrate its effectiveness over state-of-the-art interactive systems. 
Our dataset and code are publicly available.

\begin{figure*}[htb]
    \centering
    \includegraphics[width=0.9\linewidth]{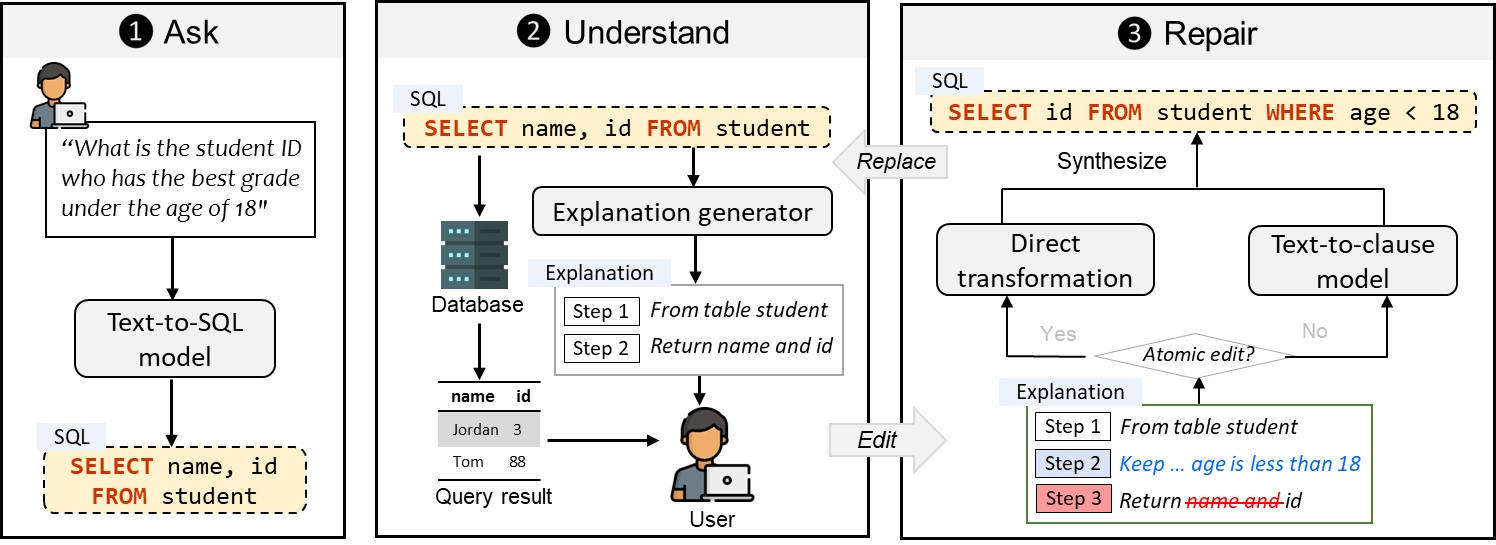}
    \caption{An Overview of Interactive SQL Generation and Refinement with Editable Step-by-Step Explanations}
    \label{fig:overview}
\end{figure*}

\section{Related Work}
\label{sec:related}

\subsection{Text-to-SQL Generation}
Natural language interfaces have long been recognized as a way to expand access to databases \citep{3}.
The construction of several large text-to-SQL datasets, such as WikiSQL \citep{wikisql} and Spider \citep{spider}, has enabled the adoption of deep learning models in this task, achieving unprecedented performance in recent years~\citep{smbop, ratsql, picard, grappa, sqlova}. Our technique is based on the recent success of neural text-to-SQL models. Unlike existing models that perform end-to-end SQL generation, we propose a new interaction mechanism for users to validate and refine generated queries through step-by-step explanations.  

As the first step to demonstrate the feasibility of our approach, we focus on single-turn SQL generation~\cite{spider} in this work. There has also been recent work that supports multi-turn SQL generation~\cite{cosql, sparc, guo2021chase}, where a sequence of interdependent queries are expressed in multiple utterances in a dialog. Models designed for multi-turn SQL generation typically need to reason about the dialog context and effectively encode the historical queries~\cite{treesql, hui2021dynamic, zhang2019editing, cai2020igsql, wang2021tracking}. Our approach can be extended to support multi-turn SQL generation by initiating separate refinement sessions for individual queries while incorporating the contextual information of previous queries into explanation generation and text-to-clause generation. 

\subsection{Interactive Semantic Parsing for SQL}
Recently, there has been a growing interest in interactive approaches that elicit user feedback to guide SQL generation. 
\citet{binary} proposed to allow users to flag incorrect queries and continuously retrain the model. 
Both DIY~\citep{diy} and NaLIR~\citep{construct_interface, nalir} enable users to select alternative values or subexpressions to fix an incorrect SQL query. PIIA~\citep{piia}, MISP~\citep{misp}, and DialSQL~\citep{dialsql} proactively ask for user feedback via multiple-choice questions. 
A common limitation of these methods is that they only solicit feedback in constrained forms, hindering their flexibility and effectiveness in addressing the variability of SQL errors.
In contrast, our approach allows more flexible feedback through direct edits to the explanations generated by the model.

The only work that supports open-ended user feedback in SQL generation is NL-EDIT \citep{nledit}. NL-EDIT is trained on SPLASH~\citep{splash}, a dataset of SQL errors and user feedback utterances.
Given an incorrect query, NL-EDIT allows users to provide a clarification utterance.
Based on the utterance, the model generates a sequence of edits to the SQL query.
Incorporating feedback expressed in a completely free-text utterance is challenging for two reasons: (1) the model needs to infer which part of the SQL query to fix; (2) the model needs to determine what changes are being requested.
In contrast, {\tool} asks users to directly edit an NL explanation and make corrections to the explanation. Comparing the initial explanation with the user-corrected explanation makes it easier to locate the part of a SQL query that needs to be changed and infer what change to make.

The idea of SQL decomposition is similar to recent work that decomposes a user question to sub-questions on SPARQL~\citep{transparent}. Their approach requires a crowd-sourced dataset to train a question decomposition model. 
In contrast, our rule-based method generates step-by-step explanations without the need for training a model. This also allows our system to map each entity in the explanation to the corresponding SQL element, making it easier for SQL correction (Sec.~\ref{Sec:text-to-clause}).



\subsection{Explaining SQL Queries in NL}

Our approach is also related to prior work that generates NL explanations for SQL queries. 
\citet{dbtalkback} argued that databases should  ``talk back'' in human language so that users can verify results. 
\citet{logos} and \citet{explaininnl} used a graph-based SQL translation approach, where each query is represented as a graph and the explanation is generated by traversing the graph. 
\citet{nledit, splash} employed a template-based explanation approach, where they manually curated 57 templates for explanation generation. These  approaches have limited capability to handle arbitrary SQL queries. To address this limitation, we propose a rule-based method to first explain terminal tokens (e.g., operators, keywords) and gradually compose them into a complete explanation based on the derivation rules in the SQL grammar. 
Another key difference is that none of the existing approaches supports {\em editable} explanations for SQL correction, which is a key feature to solicit user feedback in our approach.

\section{Approach}


Fig.~\ref{fig:overview} provides an overview of {\tool}. Given a natural language (NL) question, {\tool} invokes a text-to-SQL model to generate an initial SQL query. Then, it decomposes the generated SQL query into individual query clauses and re-orders them based on their execution order. Each clause is then translated into an NL description of the underlying data operation, which is then used to form a step-by-step explanation. By reading the NL explanation along with the query result, users can easily understand the behavior of the generated query and locate any errors, even if they are unfamiliar with SQL.

If one step is incorrect, users can directly edit its explanation to specify the correct behavior. {\tool} will then regenerate the clause based on the user-corrected explanation and update the SQL query, rather than regenerate the entire query from scratch. 
If multiple steps are incorrect, the user can add, remove, and modify all steps as needed.



\begin{figure*}[!ht]
    \centering
    \includegraphics[width=0.9\linewidth]{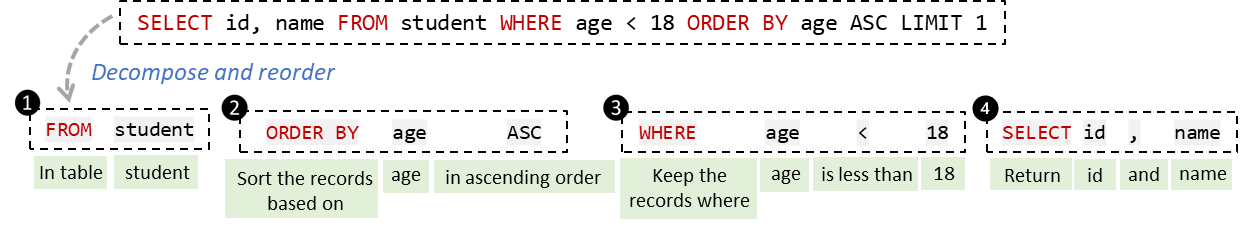}
    \caption{An example of the explanation generation process}
    \label{fig:decomposition}
\end{figure*}

\subsection{Rule-based SQL Explanation}
\label{section:explanation_generation}
To generate explanations for arbitrarily complex SQL queries (e.g., a query with nested subqueries), we design a rule-based method to first decompose a query into individual clauses. Specifically, {\tool} first parses a SQL query to its Abstract Syntax Tree (AST) based on the SQL grammar in Table~\ref{tab:table_grammar}. Then, it traverses the AST to identify the subtree of each clause while preserving their hierarchical relations.

Given the subtree of a clause, {\tool} performs an in-order traversal and translates each leaf node (i.e., a terminal token in the grammar) to the corresponding NL description based on a set of translation rules (see Table~\ref{tab:translations} in the appendices).
For example, \texttt{SELECT} is translated to ``Return'', and \texttt{Order} \texttt{By} is translated to ``Sort the records based on.'' {\tool} concatenates these descriptions to form a complete sentence as the explanation of the clause.

\begin{figure*}[htb]
    \centering
    \includegraphics[width=0.8\linewidth]{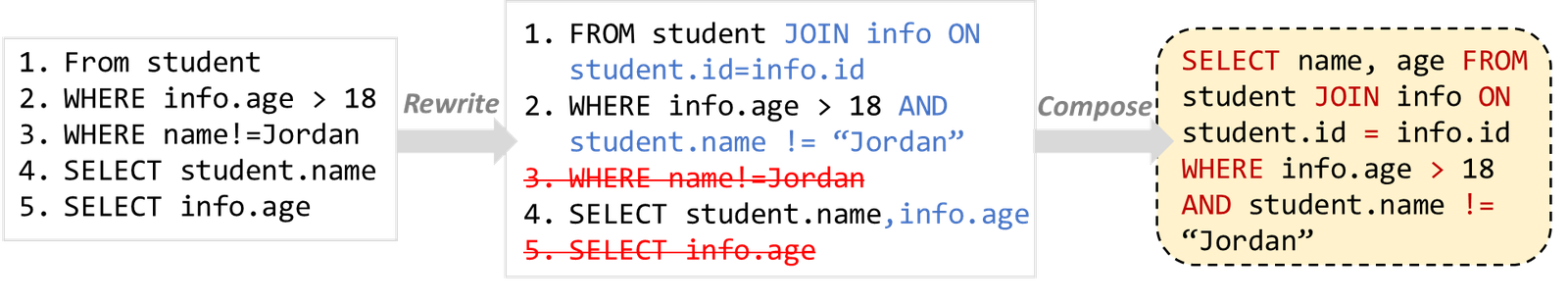}
    \caption{An example of SQL clause rewriting and composition}
    \label{fig:compose}
\end{figure*}

Since SQL engines follow a specific order to execute individual clauses in a query\footnote{\url{https://sqlbolt.com/lesson/select_queries_order_of_execution}}, {\tool} further reorders the clause explanations to reflect their execution order. We believe this is a more faithful representation of the query behavior and thus can help users better understand the underlying data operations, compared with rendering them based on the syntactic order of clauses. Fig.~\ref{fig:decomposition} shows an example translation.

\subsection{Text-to-Clause Generation}
\label{Sec:text-to-clause}

Users make edits to the explanation produced by our system to make it consistent with their goal.
Given these edits, {\tool} uses a hybrid method to generate the corresponding SQL clause.
For simple edits, such as replacing a column name, {\tool} directly edits the original clause to fix the error using three direct transformation rules (\S~\ref{Sec:direct_transformation}). For more complex edits, {\tool} uses a neural text-to-clause model to generate the clause based on the user-corrected explanation (\S~\ref{Sec:text_to_clause_model}).

The hybrid method is inspired by the findings from our recent study \cite{ning_empirical_2023}. Specifically, a large portion of SQL generation errors are simple errors (e.g., incorrect column names and operators), which can be fixed with  small edits. 
After SQL decomposition by our approach, many larger errors are further decomposed into a set of simpler errors, contained within separate clauses.
Thus, it is not necessary to regenerate the entire clause to fix such errors. 
Furthermore, compared to using a large model, direct transformation is more computationally efficient. Our experiment shows that direct transformation is 22K times faster than the text-to-clause model (Table~\ref{tab:hybrid}).

\begin{algorithm}[t]
\small
\caption{Direct transformation}
\label{algo:direct_transformation}
\KwIn{The original explanation \textit{e$_{o}$}; \\
    The new edited explanation \textit{e$_{n}$}; \\
    The original SQL clause \textit{s}; }
\KwOut{the updated SQL clause}

$C_o \gets$ \textsc{Chunk}($e_o$)  \label{line:1}

$C_n \gets$ \textsc{Chunk}($e_n$)  \label{line:2}

\ForEach {\textsc{(}c$_{o}$, c$_{n}$\textsc{)}  \textbf{in} \textsc{Align(}$C_{o}$, $C_{n}$\textsc{)} }  { \label{line:3}
        \textcolor[rgb]{0.1333,0.5451,0.1333} {// Replace} \;  \label{line:replace}
        
        \uIf{ 
                \textsc{BothColumn(}c$_{o}$, c$_{n}$\textsc{)} \label{line:4}
                \textbf{or} \\
                \textsc{BothTable(}c$_{o}$, c$_{n}$\textsc{)} \textbf{or} \\
                \textsc{BothLiteral(}c$_{o}$, c$_{n}$\textsc{)}
        } 
        {
            $s$ $\gets$ $s$.\textsc{Replace}(c$_{o}$, c$_{n}$) \; \label{line:5}
        }
     \textcolor[rgb]{0.1333,0.5451,0.1333}{// Add} \;  \label{line:6} 
     
    \uElseIf{c$_{o}$ is $\varnothing$ \textbf{and} \textsc{isColumn(}c$_{n}$\textsc{)}} 
    {
        \uIf{s.\textsc{StartWith(}{\rm \texttt{"Select"}}\textsc{)}}{
        
            $s$ $\gets$ $s$.\textsc{Append}($c_n$) 
            \label{line:9}
        }
    }

    \textcolor[rgb]{0.1333,0.5451,0.1333}{// Remove} \;  \label{line:10}
    
    \uElseIf{$c_{n}$ is $\varnothing$ \textbf{and} \textsc{isColumn(}$c_{o}$\textsc{)}}{      
            $s$ $\gets$ $s$.\textsc{Remove}($c_o$) \;  \label{line:12}
    }

        }   
    
\textbf{return} 

\end{algorithm}

\subsubsection{Direct Transformation}

\label{Sec:direct_transformation}

We define three types of {\em atomic edits} that can be directly converted into SQL edits by {\tool}: (1) replacing a column name, a table name, or a literal value (i.e., string, number), (2) adding a new column name in the explanation of a \texttt{SELECT} clause, and (3) removing a column name. 

Algorithm~\ref{algo:direct_transformation} describes our direct transformation algorithm.
After chunking the text (Lines \ref{line:1}-\ref{line:2}), {\tool} aligns and compares the chunks in the original explanation with those in the user-corrected explanation, using the \citet{needleman} algorithm (Line \ref{line:3}). This allows {\tool} to detect any replacements (Line \ref{line:replace}), additions (Line \ref{line:6}), or removals (Line \ref{line:10}) of database entities in the explanation. Based on this information, {\tool} automatically edits the corresponding SQL clause without calling a neural model (Lines \ref{line:5}, \ref{line:9}, \ref{line:12}). More details of this algorithm can be found in Appendix~\ref{appendix:Algorithm_direct_transformation}.

\subsubsection{Text-to-Clause Model}
\label{Sec:text_to_clause_model}
For more complex edits, we develop a text-to-clause model. 
We adopt the model architecture of SmBoP \citep{smbop} for this model. SmBoP is a semi-autoregressive and bottom-up transformer-based semantic parser for SQL. It decodes subtrees first and then gradually combines them to form a complete AST of the final SQL. 
To train the model, we automatically created a dataset with 83K text-clause pairs based on Spider \citep{spider}.
For each SQL query in Spider, we use the explanation generation method in Section~\ref{section:explanation_generation} to decompose the query into clauses and generate an NL explanation of each clause. To improve the diversity of NL explanations, we paraphrase the original explanations in two ways. 
First, we use a rule-based method to replace words with their synonyms (details in Table \ref{tab:table_replacement} in the appendices). 
Second, we paraphrase the explanation using an automatic paraphrasing tool: QuillBot\footnote{\url{https://quillbot.com}}. 
We train the text-to-clause model using Adam with a learning rate of $1.8e-4$ and a dropout rate of $0.1$. We perform 10-fold cross-validation and the exact set match accuracy of our text-to-clause model is 90.6\% (see Appendix~\ref{text-to-clause_model} for details). 


\subsection{SQL Rewriting and Composition}

\label{Sec:compose}


After regenerating the clauses for all user-corrected explanations, {\tool} composes them together to form a new query while avoiding syntactic errors using three rewriting rules. 

Simply combining SQL clauses may lead to syntactic errors. As shown in Fig.~\ref{fig:compose}, the regenerated clause may reference another table that does not exist in the previous query, e.g., \texttt{info} in the second clause. Thus, we design three rewriting rules to fix such errors. First, if a table is referenced but is not the table in the \texttt{FROM} clause, {\tool} rewrites the \texttt{FROM} clause to join the existing table with the new table based on the foreign key. Second, if multiple \texttt{SELECT}, \texttt{WHERE}, or \texttt{HAVING} clauses are at the same hierarchical level, {\tool} merges them into a single clause. Third, if there are multiple \texttt{ORDER BY} or \texttt{GROUP BY} clauses, {\tool} only keeps the first one. Fig.~\ref{fig:compose} shows an example of the rewriting process.

\section{Experiment}
To evaluate the performance of {\tool}, we conducted quantitative experiments on the Spider benchmark~\citep{spider} with three SOTA interactive SQL generation approaches---MISP \citep{misp}, DIY \citep{diy}, and NL-EDIT \citep{nledit}.
We also explored the impact on {\tool} of different text-to-SQL models, different task difficulties, and four different benchmarks~\citep{spider, spider-dk, spider-syn, wikisql}.
Finally, we conducted an ablation study for our hybrid method.

\subsection{Automated User Simulation \& Setup}
\label{sec:simulation}
For our quantitative evaluation of {\tool}, we developed an automated script to simulate user feedback following the user simulation method of \citet{misp}.
This setup assumes we have a user who perfectly identifies all errors and provides clear corrections.
The purpose of this experiment is to measure the upper bound of system performance without human errors.

We do this as follows. Given a generated query and the ground-truth query, our script decomposes both of them into clauses using the method described in Section~\ref{section:explanation_generation}. Then, it compares the clauses and checks their semantic equivalence using the component matching method of \citet{spider}. For example, \texttt{SELECT name, age} is considered semantically equivalent to \texttt{SELECT age, name}. 
The simulated user provides feedback when a clause in the generated query is not semantically equivalent to the corresponding clause in the ground truth (i.e., there is an error).

We simulated three types of mismatches.
First, if the generated query contains a clause that does not exist in the ground truth, our script will delete its explanation from the original explanation. Second, if the generated query is missing a clause from the ground truth, our script will generate the explanation of this missing clause using the explanation generation method described in Section~\ref{section:explanation_generation}, paraphrase it using QuillBot, and insert it into the corresponding location of the original explanation. Finally, if the generated query contains a clause that is inconsistent with the ground truth, our script will generate the explanation based on the correct clause in the ground truth, paraphrase it using QuillBot, and replace the explanation of the incorrect clause with the paraphrased one.





\subsection{Comparison Systems}
\label{sec:baselines}

We compared {\tool} to three state-of-the-art interactive SQL generation methods:

Among tools that allow users to give feedback by answering multiple-choice questions~\citep{dialsql, piia, misp}, we select MISP~\citep{misp} to compare with because it has the best performance in simulation. During the interaction,  MISP asks users to clarify whether a column should be considered in the query, and the user can answer yes or no. The user's answer is used to constrain the decoding process by adjusting the probability of code tokens induced by the answer.
We used the original implementation of MISP from their GitHub repository. Furthermore, since their GitHub repository provides a user simulation script, we reuse it for the user simulation in our experiments. 

DIY \citep{diy} enables users to refine a generated SQL query by showing the table names, column names, operators, and aggregate functions that correspond to words in the NL question and allowing the user to select alternatives from drop-down menus.
We reimplemented DIY since no open-source implementation is available.
We cannot directly compare with the user performance from the DIY paper because they did not report any objective measures, such as task completion rates and time \citep{diy}.
To construct the word-entity mapping in DIY, we calculate word embedding semantic similarity. In the user simulation, we align the generated SQL with the ground truth SQL. If an entity in the generated SQL is not present in the ground truth SQL, which indicates an error, and it has been mapped to the NL question, which means users can give feedback via a drop-down menu, we replace it with the corresponding ground truth entity.

NL-EDIT \citep{nledit} enables users to correct errors by giving feedback in natural language. User feedback is parsed into a set of simple edits (e.g., add, remove) that are applied to the SQL query.
We worked with the NL-EDIT authors to run their system, but were unable to resolve issues due to missing code and other run-time errors.
We report results for NL-EDIT using the accuracy numbers from the NL-EDIT paper.




\begin{table}
\centering
\resizebox{0.417\textwidth}{!}{%
\begin{tabular}{@{}lc@{}}
\toprule
  & \textbf{Acc$_{\scriptsize \rm  \textbf{set}}$}
   \\ 
  \midrule
  
EditSQL \citep{zhang2019editing}        & 0.576                                             \\
\specialrule{0.02em}{2pt}{2pt}
\textbf{Human-in-the-Loop Methods} \\
+ MISP \citep{misp}        & 0.644                                                \\
+ DIY \citep{diy}    & 0.647                                               \\
+ NL-EDIT \citep{nledit}    & 0.666                                              \\
+ STEPS (Ours)      & \textbf{0.979}                                               \\ 
\midrule\midrule
\textbf{AI-Only Methods} \\
Graphix-3B + PICARD~\citep{graphix-t5}       & 0.771                                               \\ 
SHiP + PICARD~\citep{ship}        & 0.772                                               \\ 
RESDSQL-3B + NatSQL~\cite{resdsql}  & 0.805                               \\
\bottomrule
\end{tabular}}%
\caption{Exact Set Matching Accuracy Comparison. Note, these results are on the dev set as we are unable to use the hidden test set in the human experiments.} 
\label{tab:result2}
\end{table}

\subsection{Results}



\begin{table*}
\centering
\small
\resizebox{0.95\textwidth}{!}{
\begin{tabular}{c|ccccc|ccccc}
\toprule
& \multicolumn{5}{|c|}{\textbf{Acc}$_{\scriptsize \rm  \textbf{set}}$} & \multicolumn{5}{c}{\textbf{Acc}$_{\scriptsize \rm  \textbf{exec}}$} \\
   & \textbf{Easy} & \textbf{Medium} & \textbf{Hard} & \textbf{Extra hard} & \textbf{All} & \textbf{Easy} & \textbf{Medium} & \textbf{Hard} & \textbf{Extra hard} & \textbf{All} \\
\midrule
EditSQL  & 0.681 & 0.632 & 0.456 & 0.395 & 0.576 & -     & -     & -     & -     & -     \\ 
+ STEPS  & 0.991 & 1.000 & 0.976 & 0.912 & 0.979 & 0.991 & 0.995 & 0.939 & 0.912 & 0.971 \\
\midrule
SmBoP    & 0.883 & 0.791 & 0.655 & 0.512 & 0.745 & 0.718 & 0.669 & 0.672 & 0.518 & 0.657 \\ 
+ STEPS  & 0.992 & 1.000 & 0.977 & 0.916 & 0.981 & 0.992 & 0.995 & 0.943 & 0.916 & 0.973 \\
\bottomrule
\end{tabular}}
\caption{{\tool}'s Accuracy on SQL Tasks with Different Levels of Difficulty}
\label{tab:breakdown_accuracy}
\end{table*}


\begin{table}
\centering
\small
\setlength{\tabcolsep}{4pt}
\begin{tabular}{lcccc} 
\toprule
& \multicolumn{2}{c}{\textbf{Acc$_{\scriptsize \rm  \textbf{set}}$}} & \multicolumn{2}{c}{\textbf{Acc$_{\scriptsize \rm  \textbf{exec}}$}} \\
\midrule
& SmBoP & + \tool & SmBoP & + \tool \\
\midrule
WikiSQL              &  0.862  &  0.983   & 0.895    & 0.980             \\
Spider   & 0.745    & 0.981  & 0.657    & 0.973         \\
Spider-DK       &  0.534  & 0.987  & 0.537    & 0.976                \\
Spider-Syn              &  0.572  &  0.969  & 0.600    & 0.972             \\
\bottomrule
\end{tabular}
\caption{Evaluation on different datasets}
\label{tab:eval_datasets}
\end{table}

\begin{table}
\centering
\small
\setlength{\tabcolsep}{4pt}
\begin{tabular}{lccc} 
\toprule
                      & \textbf{Acc$_{\scriptsize \rm  \textbf{set}}$}& \textbf{Acc$_{\scriptsize \rm  \textbf{exec}}$} & \textbf{Time (ms)}  \\

\midrule
Direct transform only &0.788   & 0.745    & 0.0042         \\
Text-to-clause only       &  0.981  & 0.973    &  95.53            \\
Hybrid              &  0.981  &  0.973   &  57.24            \\
\bottomrule
\end{tabular}
\caption{Ablation Study of the Hybrid Method}
\label{tab:hybrid}
\end{table}

\textbf{Comparison with the Three SOTA Interactive Approaches.} 
Table \ref{tab:result2} shows the exact set match accuracy of {\tool}, MISP, DIY, and NL-EDIT. Following the experimental design of MISP and NL-EDIT, we use EditSQL \citep{zhang2019editing} as the base SQL generation model and exact set matching accuracy \citep{spider} as the evaluation metric. {\tool} achieves 97.9\% accuracy, outperforming all three previous approaches by at least 31\%.

\textbf{Comparison with Strong Text-to-SQL Models.} 
Table~\ref{tab:result2} also shows the exact set match accuracy of three high-performing text-to-SQL models~\citep{graphix-t5, ship, resdsql}.\footnote{As the test set of Spider is not released, we selected the top three models based on their exact set match accuracy on the development set at the time of our experiments.}
Compared with these models, {\tool} achieved 17\%-20\% accuracy improvement by soliciting user feedback. This indicates that allowing users to edit step-by-step explanations can produce results that are far better than the best pure-AI models while also providing users with confidence that the query is doing what they want.



\textbf{Evaluation with Different Base Models \& Task Difficulty Levels.}
To demonstrate {\tool}'s performance is generalizable to other base models, we also evaluate {\tool} on another model called SmBoP \citep{smbop}. 
SmBoP is one of the best models on the Spider leaderboard with $74.5\%$ exact set matching accuracy.
Table~\ref{tab:breakdown_accuracy} shows {\tool}'s exact set matching accuracy with SmBoP as the base model in comparison to EditSQL. We also report execution accuracy, another popular metric that compares the query results between the generated query and the ground truth. Note that since EditSQL does not predict any value in SQL conditions, the queries generated by EditSQL are not runnable. Thus, we cannot measure the execution accuracy of EditSQL. The result shows that {\tool} consistently improves the accuracy of both models on SQL tasks with various levels of difficulty.\footnote{Spider categorizes their SQL tasks into four difficulty levels---easy, medium, hard, and extra hard.} Specifically, {\tool} can almost solve all easy and medium tasks and also achieves more than 90\% accuracy for the hard and extra hard tasks.
For hard and extra hard tasks, the generated SQL queries often include more errors.
It can be challenging for other approaches to fix all of them at once. In our case, decomposing the original task into smaller steps makes fixing multiple errors as easy as fixing one.





\textbf{Generalizability to Different Datasets.}
To further demonstrate the generalizability of {\tool}, we evaluate {\tool} on three other datasets——Spider-DK~\citep{spider-dk}, Spider-Syn~\citep{spider-syn}, and WikiSQL~\citep{wikisql}. 
Note that since {\tool} is trained on Spider, its models are out-of-domain when applying to different datasets. Table~\ref{tab:eval_datasets} demonstrates that \tool achieves comparable performance across datasets.

\textbf{Ablation Study for the Hybrid Method.}
Table~\ref{tab:hybrid} shows the ablation results of the hybrid method of {\tool}. Regarding SQL generation accuracy, {\tool} achieves comparable accuracy when using text-to-clause alone, while experiencing a significant accuracy degradation when using direct transformation alone. This makes sense since the direct transformation method is only designed to fix a subset of the possible error types.
However, for the types for which it is intended, the direct transformation approach is very accurate.
As a result, using it as part of the hybrid system increases efficiency without decreasing accuracy.


\section{User Study}
\label{sec:user}

To evaluate the usability and accuracy of {\tool} when interacting with real users, we conducted a within-subjects user study with 24 participants.\footnote{Our study was approved by Purdue University's IRB.}

\subsection{Participants}
We recruited 24 participants (22M, 2F) through mailing lists at Purdue University. 
In the recruitment email, we shared a consent form that included detailed information about the study procedure, potential risks, data usage, and confidentiality. We obtained consent from each user before proceeding with the study. All collected data were anonymized and de-identified.
Each participant received a \$25 gift card as compensation for their time.

To investigate how user expertise affects the performance of {\tool}, participants were selected based on their familiarity with SQL. Specifically, 10 of them had never heard about or used SQL before (end-user); 10 knew the basics of SQL but had to search online to recall the syntactic details when writing a SQL query (novice); 4 could fluently write SQL queries (expert).

\subsection{Comparison Systems}
We used MISP \citep{misp} and DIY \citep{diy} as comparison systems. As explained in Section~\ref{sec:baselines}, we did not use NL-EDIT, since we were unable to reproduce it.
To ensure a fair comparison, we developed user interfaces with the same visual style for {\tool}, MISP, and DIY. 

\subsection{SQL Tasks \& Procedures}


Each study includes 3 sessions, one for each tool. In each session, participants were asked to use the assigned tool to complete 8 SQL tasks in 10 minutes. The time limit was decided by 4 pilot studies, allowing sufficient time to  complete multiple tasks. 
To select the tasks, we first performed stratified random sampling on Spider to create a task pool of 24 SQL tasks, including 6 easy tasks, 6 medium tasks, 6 hard tasks, and 6 extra hard tasks. Before each session, we selected 2 tasks from each difficulty level in the task pool, which constitutes a total of 8 tasks to be solved in the session. 
To mitigate learning effects, the orders of both task assignment and tool assignment order were counterbalanced across participants.

Each session started with participants watching a tutorial video of the assigned tool (6 min for {\tool}, 3 min for MISP, and 2 min for DIY). 
The {\tool} video was longer simply because {\tool} had more features. During all tutorials, we allowed users to pause the video and ask questions.
Participants were then given 5 minutes to practice and get familiar with the tool before working on real tasks.  
For each task, participants were asked to read the description of the task and then ask an initial NL question to the assigned tool. After receiving the generated query and results, participants could validate and repair the generated query using the tool. Participants were allowed to skip a task if they found it too hard to solve.
Participants could view the produced SQL, but were not given the ability to edit the SQL directly.

At the end of each session, participants were asked to complete a post-task survey to rate their confidence about the final SQL query, how successful they perceived themselves in completing the tasks, and the mental effort to complete the tasks on a 7-point Likert scale. After all three sessions, participants completed a final survey, in which they directly compared the three tools. We recorded each study with the permission of the participants. Each study took an average of 79 minutes.

\begin{table}
\centering
\resizebox{0.43\textwidth}{!}{%
\begin{tabular}{lcccc}
\toprule
  & \textbf{Complete}
  & \textbf{Correct} 
  & $\rm \textbf{Acc.}$
  & \textbf{Skipped} \\ 
  \midrule

MISP        & 3.0~~    & 1.7~~      & 0.57~~    & 1.4~~   \\
DIY     & 5.4~~    &  3.5~~     &  0.68~~     & 0.8~~  \\
STEPS       & \textbf{6.7$\uparrow$}   & \textbf{5.7$\uparrow$}     & \textbf{0.86$\uparrow$}  & \textbf{0.3$\downarrow$}\\ 
\bottomrule
\end{tabular}%
}
\caption{User Performance (best results in bold). For all metrics, an ANOVA test indicated statistically significant mean differences across 3 tools ($p$-value < 0.01).}
\label{tab:task_completion}
\end{table}

\subsection{Results}



\textbf{Task Completion Rate Analysis.}
Table \ref{tab:task_completion} shows the average number of completed tasks, correct completions, task completion accuracy (\#correct / \#completed), and skipped tasks. We found that participants using {\tool} completed more tasks compared to those using MISP and DIY. Furthermore, participants using {\tool} completed significantly more tasks correctly than DIY and MISP, achieving the highest accuracy (85.81\%) in SQL generation. Participants using {\tool} barely skipped a task, implying that {\tool} provided sufficient support for users to tackle challenging tasks so that users did not give up quickly. The ANOVA test indicates that the mean differences in Table~\ref{tab:task_completion} are statistically significant among all three conditions ($p$-value < 0.01). These results indicate that {\tool} can help users complete SQL tasks more efficiently and correctly than prior methods. 

    
    

\begin{figure}
    \includegraphics[width=0.95\linewidth]{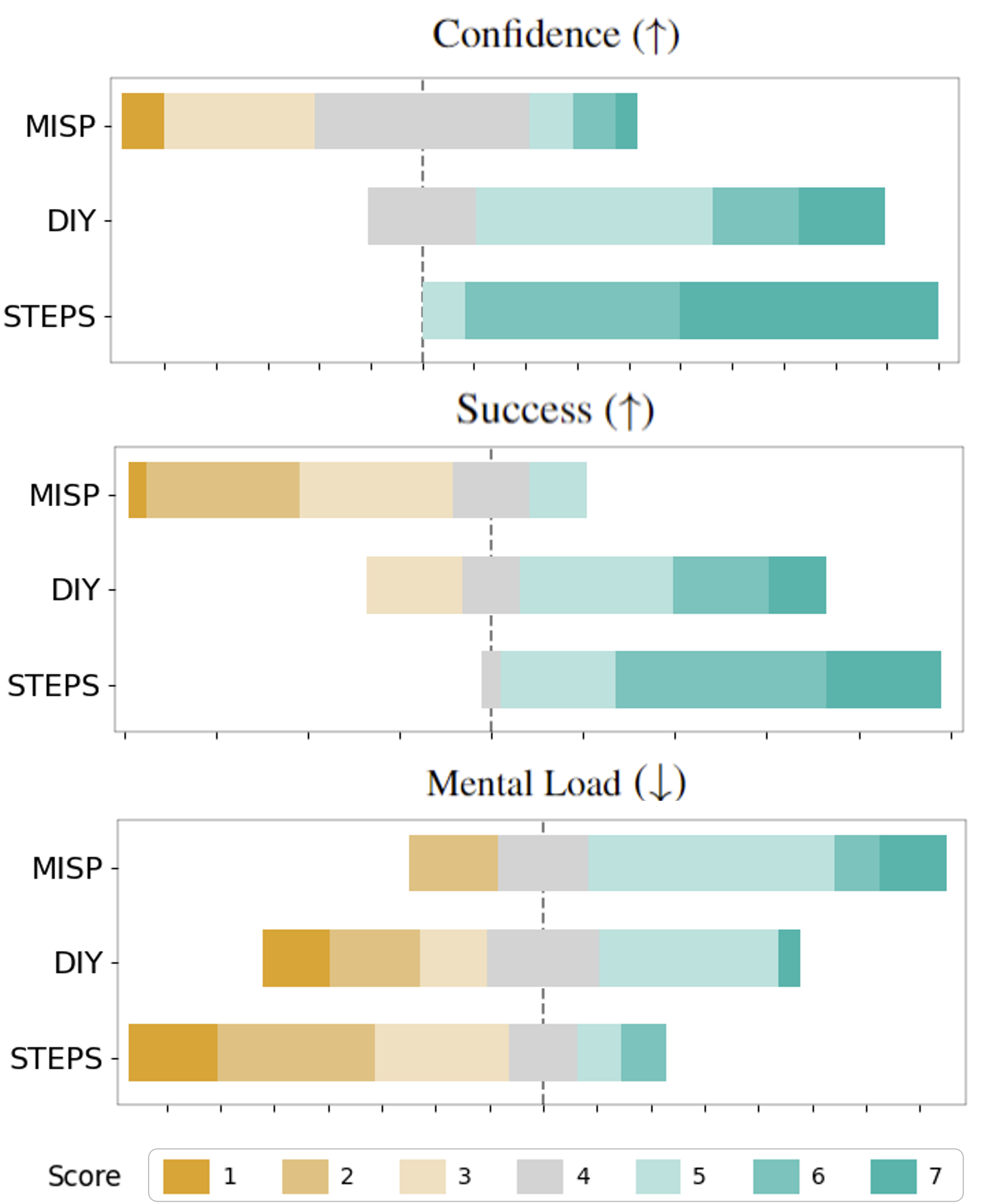}
    \caption{User Perception.}
    \label{fig:perception}
\end{figure}

\textbf{The Impact of SQL Expertise.}
We further investigated whether the SQL expertise of users has an impact on user performance. 
We found that all three user groups performed similarly in each condition. This implies that SQL expertise does not have a significant impact on user performance when interacting with STEPS. 
For a visualization of the results, see Fig.~\ref{fig:expertise} in the appendices.

\textbf{User Perception.}
In the post-study survey, all participants ranked {\tool} as the most usable and useful tool. As shown in Figure \ref{fig:perception}, participants felt the most confident and successful while experiencing the least mental load when using {\tool}. 


\section{Analysis of Post-study Survey Responses}
\label{app:user}

We analyzed the post-task survey responses and interview recordings to understand why participants performed much better when using {\tool} compared with using MISP and DIY. 
17 participants strongly agreed that seeing the natural language explanation helped them understand the SQL query, and 22 participants explicitly wrote that they highly appreciated the step-by-step explanations provided by {\tool}, since these explanations made SQL queries more understandable, editable, and learnable. 
P12 wrote, ``\textit{I liked that it shows the steps in human language so if there is a mistake I can edit it easily. Also, it was nice to see the generated SQL code I believe I could learn SQL using this tool also.}'' 
In contrast, 14 of 24 participants reported it was hard to understand and validate the generated SQL queries when using MISP or DIY. 
P1 wrote, ``\textit{Sometimes it generates very complex SQL that is difficult to read and check.}''
P9 wrote, ``\textit{Sometimes it gives the wrong answer. As I'm no expert in SQL, I couldn't tell instantly if the queries were wrong, so I had to go back to the data and check manually.}''
12 participants reported that the feedback elicitation mechanism in MISP was not very efficient.
P16 wrote, 
``\textit{I have to keep answering yes or no questions when using MISP.}''
11 of them reported the drop-down menus provided by DIY limited their ability to make changes. 
P3 said, ``\textit{[It is] hard to know how to make changes / resolve issues that were not covered by the drop-down menus.}''



\section{Discussion}

Both the quantitative experiments and the user study demonstrate {\tool} can significantly improve the accuracy of SQL generation.
This is largely attributed to the interaction design, which allows users to precisely pinpoint which part of the SQL is wrong and only regenerates the incorrect clauses rather than the entire SQL query. 
In contrast, existing approaches do not support expressive ease or error isolation. Users either cannot regenerate new content (e.g., DIY), or can only regenerate the entire query rather than just the erroneous part (e.g., MISP).
\citet{ning_empirical_2023} showed that this lack of error isolation often introduces new errors, which frustrates users and makes errors hard to fix.

\textbf{Error Analysis.}
While simple errors are prevalent in SQL generation, our ablation study (Table~\ref{tab:hybrid}) shows that only fixing simple errors is insufficient, which motivates the design of our hybrid method.
Our hybrid method can handle a broad range of errors because users can flexibly correct entities or clauses in a query. This ability helps reduce the difficulty of tasks by dividing complex errors into simpler ones, allowing users to solve them separately.

In our automated user simulation, {\tool} failed in a few cases when the text-to-clause model predicted the wrong clause type. For example, the paraphrased ground truth explanation of one step was: ``\textit{Ensure that all categories where the total cost of therapy exceeds 1000 are included.}'' The text-to-clause model predicted a  \texttt{WHERE} clause instead of a \texttt{HAVING} clause.

In the user study, one common challenge arose when multiple tables in the database had the same column name.
If users did not look carefully at the database schema, they may have not explicitly indicated the table to be used.
That creates an ambiguity for the model.

\textbf{Other Datasets and Domains.}
Our system should work for any SQL dataset, as our approach is domain-agnostic and covers general SQL structures.
For other forms of code,  such as WebAPI \citep{webapi} and SPARQL \citep{SPARQL, transparent}, the general idea is applicable, but new models would be needed for (a) code generation, (b) explanation generation, and (c) code correction.

\section{Conclusion}

This work presents {\tool}, a new interactive approach for text-to-SQL generation. {\tool} decomposes a text-to-SQL task into smaller text-to-clause tasks and enables users to validate and refine a generated query via editable explanations. Experiments on four benchmarks and a user study show {\tool} can significantly boost the accuracy of end-to-end models by incorporating user feedback. {\tool} significantly outperforms three state-of-the-art approaches for interactive SQL generation across all metrics considered.


\section{Limitations}


Our automated user simulation is an optimistic experiment that does not account for user errors, such as not being able to identify mistakes in the explanation. 
The simulation was designed to test a scenario in which a user can perfectly identify which step of the explanation is wrong and accurately describe a corrected version in natural language. Creating such a perfect user required the use of the ground truth, both for the identification step and to generate the natural language correction. This simulation is not representative of real-world use.
That limitaton was the motivation for our study with real users, in which we had actual people use different tools without information about correct answers. As shown in Table~\ref{tab:task_completion}, the accuracy of the user study is lower than the simulation, but {\tool} is still very effective and outperforms other tools.
We choose to include the simulation study because it shows the potential for {\tool} to make corrections if there is no human error. 

In this paper, we only evaluate {\tool} on single-turn SQL generation. In future work, our approach can be extended to multi-turn SQL generation by incorporating contextual information when editing the natural language explanation.

While our approach is designed to be general for SQL generation and potentially other code generation tasks, the current version only supports SQL keywords that appear in the Spider dataset.
Like other text-to-SQL datasets, Spider only covers query operations (e.g., \texttt{SELECT}) and does not cover update operations (e.g., \texttt{INSERT}) for evaluation convenience. 
But it would be straightforward to cover unsupported operations by adding new translation rules.

\section{Ethical Consideration}
The interactive text-to-SQL system proposed by this work poses minimal risks to human users and society. Instead, it will significantly lower the barrier of querying database systems and empower a great number of people, especially those without technical backgrounds, to access and analyze data. 
To evaluate the usability of our system, we conducted a human-subject study with real users. To minimize the risks to human subjects, we strictly followed the community standards with the approval from the Purdue University IRB office.
In the recruitment email, we shared a consent form that includes detailed information about the study procedure, potential risks, data usage, and confidentiality. We obtained consent from each user before proceeding with the study. All collected data were anonymized and de-identified to protect the privacy of users.

\section*{Acknowledgments}

This material is based in part on work supported by an Amazon Research Award, the Australian Research Council through a Discovery Early Career Researcher Award and by the Defense Advanced Research Projects Agency (grant \#HR00112290056).


\bibliography{anthology,custom}
\bibliographystyle{acl_natbib}

\appendix

\section{SQL Grammar and Translation Rules}
\label{app:grammar}

\begin{table}[!htb]
\resizebox{0.42\textwidth}{!}{
\centering
\begin{tabular}{l} 
\toprule

\begin{tabular}[c]{@{}l@{}} $\langle$ sql $\rangle$  := \texttt{SELECT} \ $\langle$ nouns $\rangle$\ $\langle$ sub $\rangle$\  \\ \qquad\quad\textbar{} $\langle$ sql $\rangle$\  \texttt{INTERSECT} \ $\langle$ sql $\rangle$\ \\ \quad\qquad\textbar{} $\langle$ sql $\rangle$\ \texttt{UNION} \ $\langle$ sql $\rangle$\ \\ \qquad\quad\textbar{} $\langle$ sql $\rangle$\ \texttt{EXCEPT} \ $\langle$ sql $\rangle$\ \end{tabular}                                                                                                              \\ 
\specialrule{0.002em}{0.5pt}{0.5pt}
\begin{tabular}[c]{@{}l@{}} $\langle$ sub  $\rangle$ := \ $\epsilon$\\  
\qquad\quad\textbar{} \texttt{FROM} $\langle$ noun $\rangle$\ $\langle$ sub $\rangle$\ \\
\qquad\quad\textbar{} \texttt{WHERE} $\langle$ condition $\rangle$\  $\langle$ sub $\rangle$\ 
 \\\qquad\quad\textbar{} \texttt{JOIN} $\langle$ noun $\rangle$\  \texttt{ON} \  $\langle$ condition $\rangle$\   $\langle$ sub $\rangle$ \\ \qquad\quad\textbar{} \texttt{GROUP}~\texttt{BY} $\langle$ noun $\rangle$\ $\langle$ sub $\rangle$\ \\ \qquad\quad\textbar{} \texttt{HAVING} $\langle$ condition $\rangle$\  $\langle$ sub $\rangle$\  \\ \qquad\quad\textbar{} \texttt{ORDER}~\texttt{BY} $\langle$ noun $\rangle$\  $\langle$ sorting $\rangle$\  $\langle$ sub $\rangle$\ \\ \qquad\quad\textbar{} \texttt{LIMIT} \texttt{NUM} \end{tabular}  \\ 
\specialrule{0.002em}{0.5pt}{0.5pt}
\begin{tabular}[c]{@{}l@{}} $\langle$ nouns  $\rangle$ := \texttt{DISTINCT} $\langle$ nouns $\rangle$  \\ \qquad\qquad \textbar{} $\langle$ noun $\rangle$,$\langle$ nouns $\rangle$ \\ \qquad\qquad \textbar{} $\langle$ noun $\rangle$ \\ \qquad\qquad \textbar{} $\langle$ func $\rangle$ {\rm (} $\langle$ noun $\rangle$ {\rm )} \end{tabular}                                                                                                                                                                                 \\ 
\specialrule{0.002em}{0.5pt}{0.5pt}
\begin{tabular}[c]{@{}l@{}}$\langle$ condition $\rangle$ := $\langle$ noun $\rangle$\  $\langle$ op $\rangle$\ \texttt{NUM} \\ \qquad\qquad\quad \textbar{} $\langle$ noun $\rangle$\  $\langle$ op $\rangle$\  $\langle$ noun $\rangle$ \\\qquad\qquad\quad\textbar{} $\langle$ noun $\rangle$\  $\langle$ op $\rangle$\  $\langle$ sql $\rangle$\ \\ \qquad\qquad\quad\textbar{} \texttt{BETWEEN} $\langle$ noun $\rangle$\  \texttt{AND} \ $\langle$ noun $\rangle$\  \\ \qquad\qquad\quad\textbar{} $\langle$ condition $\rangle$\  \texttt{AND}\  $\langle$ condition $\rangle$\  \\ \qquad\qquad\quad\textbar{} $\langle$ condition $\rangle$\  \texttt{OR}\  $\langle$ condition $\rangle$\  \\ \qquad\qquad\quad\textbar{} \texttt{NOT} $\langle$ condition $\rangle$ \end{tabular}                          \\ 
\specialrule{0.002em}{0.5pt}{0.5pt}
$\langle$ sorting $\rangle$ \rm := \texttt{ASC} \textbar{} \texttt{DESC} \textbar{} $\epsilon$                                                                                                                                                                                                     \\ 
\specialrule{0.002em}{0.5pt}{0.5pt}
$\langle$ func $\rangle$ := \texttt{COUNT} \textbar{} \texttt{AVG} \textbar{} \texttt{MAX} \textbar{} \texttt{MIN} \textbar{} \texttt{SUM}                                                                                                                                                                                                 \\ 
\specialrule{0.002em}{0.5pt}{0.5pt}
$\langle$ op $\rangle$ := \textgreater{}= \textbar{} \textless{}= \textbar{} \textgreater{} \textbar{} \textless{} \textbar{} = \textbar{} !=                                                                                                                                        \\ 
\specialrule{0.002em}{0.5pt}{0.5pt}
$\langle$ noun $\rangle$ := \texttt{STRING} \textbar{} \texttt{STRING}.\texttt{STRING} \textbar{} *                                                                                                                                                                                                                                \\ 
\bottomrule
\end{tabular}
}
\arrayrulecolor{black}
\caption{A Simplified SQL Grammar}
\label{tab:table_grammar}
\end{table}

Table~\ref{tab:table_grammar} shows a simplified version of the SQL grammar. In this grammar, italicized text with angle brackets, such as $\langle sql \rangle$, represents non-terminals which can be further expanded based on derivation rules. Text without brackets, such as the \texttt{SELECT} keyword, represents terminals that cannot be further expanded. Using the derivation rules in Table~\ref{tab:table_grammar}, {\tool} decomposes a SQL query into 6 types of SQL clauses: (1) \texttt{FROM-JOIN-ON}, (2) \texttt{WHERE}, (3) \texttt{GROUP BY}, (4) \texttt{HAVING}, (5) \texttt{ORDER BY}, (6) \texttt{SELECT}. We do not separate the \texttt{JOIN} clause from the \texttt{FROM} clause, since it is easier to translate them together. For nested queries with \texttt{INTERSECT}, \texttt{UNION}, \texttt{EXCEPT}, \texttt{NOT} \texttt{IN} keywords, {\tool} first decomposes them into subqueries and then decompose each subquery to the 6 types of clauses above.

{\tool} translates each SQL clause to a natural language explanation based on translation rules and templates in Table~\ref{tab:translations} and Table~\ref{tab:explanation2}.  Table~\ref{tab:translations} shows the translation rules for individual SQL tokens, e.g., keywords, operators, built-in functions, etc. Specifically, \{col\} and \{T\} mean translating a column or table name to a more readable name. 
We pre-defined mapping between each table and column in a database to a more readable name. 
Such a mapping can be easily defined based on the database schema and only needs to be defined once. If no such mapping is available, {\tool} will reuse the same column/table name as defined in the database schema. Table~\ref{tab:explanation2} shows the translation templates for nested queries. The \textsc{Translate} function means recursively invoking the explanation generation method on the subquery.

\begin{table}[t]
\resizebox{0.46\textwidth}{!}{
    \centering
    \begin{tabular}{ll}
    \hline
    \textbf{SQL Elements} & \textbf{Translation}\\
    \hline
    \texttt{SELECT} & Return \\ 
    \texttt{FROM} & In table \\ 
    \texttt{JOIN} & and table \\ 
    \texttt{WHERE} & Keep the records where \\ 
    \texttt{GROUP BY} & Group the records based on \\ 
    \texttt{HAVING} & Keep the groups where \\ 
    \texttt{ORDER BY} & Sort the records based on \\ 
    \texttt{LIMIT 1} & return the first record \\
    \texttt{LIMIT num} & return the top \texttt{{num}} records \\
    \texttt{*} & all the records \\
    $\texttt{col}_1, \texttt{col}_2$ & the $\texttt{\{col}_1$\} and the $\texttt{\{col}_2$\} \\
    $\texttt{c}_1\, \texttt{c}_2\, \texttt{c}_3$ &  the $\texttt{\{c}_1$\}, the $\texttt{\{c}_2$\} and the $\texttt{\{c}_3$\}\\
    \texttt{T.col} & \texttt{\{col\}} of \texttt{\{T\}} \\ 
    \texttt{COUNT(col)} & the number of \texttt{\{col\}} \\ 
    \texttt{COUNT(*)} & the number of records \\ 
    \texttt{AVG(col)} & the average value of \texttt{\{col\}} \\ 
    \texttt{MAX(col)} & the maximum value of \texttt{\{col\}} \\ 
    \texttt{MIN(col)} & the minimum value of \texttt{\{col\}} \\ 
    \texttt{SUM(col)} & the sum value of \texttt{\{col\}} \\ 
    \texttt{ASC} & in ascending order \\ 
    \texttt{DESC} & in descending order \\ 
    \texttt{=} & is \\ 
    \texttt{!=} & is not \\ 
    \texttt{>} & is greater than \\ 
    \texttt{>=} & is greater than or equal to \\ 
    \texttt{<} & is less than \\ 
    \texttt{<=} & is less than or equal to \\ 
    \texttt{IN} & is in \\ 
    \texttt{NOT IN} & is not in \\ 
    \texttt{BETWEEN} & is between \\ 
    \texttt{LIKE} & is in the form of \\ 
    \texttt{NOT LIKE} & is not in the form of \\ 
    \hline
    \end{tabular}
}
\caption{Translation rules for SQL elements}
\label{tab:translations}
\end{table}

\begin{table}
\resizebox{0.51\textwidth}{!}{
\centering

\begin{tabular}{ll}
\hline
\textbf{SQL compound} & \textbf{Translation}\\
\hline
\specialrule{0em}{2pt}{2pt}
\multirow{5}{*}{\texttt{q$_1$ INTERSECT q$_2$}}  & Start the first query: \\ & \textsc{Translate}(q$_1$); \\
                                & Start the second query; \\ & \textsc{Translate}(q$_2$); \\
                                & Return the intersection of them; \\
\specialrule{0.002em}{2pt}{3.5pt}

\multirow{5}{*}{\texttt{q$_1$ UNION q$_2$}}  & Start the first query {q$_1$}: \\ & \textsc{Translate}(q$_1$); \\
                                & Start the second query: \\ & \textsc{Translate}(q$_2$); \\
                                & Return the union of them. \\
\specialrule{0.002em}{2pt}{3.5pt}

\multirow{5}{*}{\texttt{q$_1$ EXCEPT q$_2$}}   & Start the first query: \\ & \textsc{Translate}(q$_1$); \\
                                & Start the second query: \\
                                & \textsc{Translate}(q$_2$); \\
                                & Return the records in {q$_1$} but not in {q$_2$}. \\

\specialrule{0.002em}{2pt}{3.5pt}

\multirow{5}{*}{... \texttt{col IN/NOT IN q$_1$}}   & Start the first query: \\
& \textsc{Translate}(q$_1$); \\
                                & Start the second query: \\
                                & \textsc{Translate}(...); \\
                                & Keep the records where \{col\} in/not in {q$_1$}. \\

\specialrule{0em}{2pt}{2pt}
\hline
\end{tabular}
}
\caption{NL explanation translation rules for SQL compound}
\label{tab:explanation2}
\end{table}

\begin{table*}[h]
\centering
\resizebox{0.7\textwidth}{!}{%
\begin{tabular}{cl}
\toprule
\multicolumn{1}{c}{\textbf{Template word}} &
  \multicolumn{1}{c}{\textbf{Substitute synonyms}} \\ 
\toprule
\specialrule{0em}{1pt}{1pt}
return &
  \begin{tabular}[c]{@{}l@{}}get, find, find out, discover,   show, show me, determine, \\ demonstrate, give me, obtain, select, choose,   search, \\ choose, search, display, list, acquire, gain\end{tabular} \\ \specialrule{0.01em}{2pt}{2pt}
keep the   records where & make, make sure, where, filter the records where                                                                                       \\ \specialrule{0.01em}{2pt}{2pt}
greater   than &
  \begin{tabular}[c]{@{}l@{}}more than, exceed, no less than,  over, above, \\ larger than, beyond, in excess of, transcend, surpass\end{tabular} \\ \specialrule{0.01em}{2pt}{2pt}
less than                & \begin{tabular}[c]{@{}l@{}}lower than, no more than, below,   lesser, under, \\ underneath, not so much as, beneath\end{tabular}      \\ \specialrule{0.01em}{2pt}{2pt}
ascending                & \begin{tabular}[c]{@{}l@{}}increasing, ascendant, growing, rising, \\ soaring, climbing, mounting\end{tabular}                         \\ \specialrule{0.01em}{2pt}{2pt}
descending               & \begin{tabular}[c]{@{}l@{}}decreasing, descendant, falling, declining, \\ dropping, lessening, diminishing\end{tabular}                \\ \specialrule{0.01em}{2pt}{2pt}
maximum                  & \begin{tabular}[c]{@{}l@{}}max, maximum, utmost, greatest, \\ most, topmost, highest, top, largest, biggest\end{tabular}               \\ \specialrule{0.01em}{2pt}{2pt}
minimum                  & \begin{tabular}[c]{@{}l@{}}lowest, smallest, least, min, minimal,\\ bottom, bottommost, lowermost\end{tabular}                         \\ \specialrule{0.01em}{2pt}{2pt}
number of                & amount of, quantity of, total of                                                                                                       \\ \specialrule{0.01em}{2pt}{2pt}
in the  form of          & appearing as, with the appearance of, in the shape of                                                                                  \\ \specialrule{0.01em}{2pt}{2pt}
that has                 & associated with, connected to                                                                                                          \\ \specialrule{0.01em}{2pt}{2pt}
based on                 & \begin{tabular}[c]{@{}l@{}}according to, in terms of,   specified by, \\ built on, established on, considering, regarding\end{tabular} \\ \specialrule{0.01em}{2pt}{2pt}
distinct                 & \begin{tabular}[c]{@{}l@{}}different, disparate,   distinctive, \\ particular, diverse, dissimilar, unique\end{tabular}                \\ \specialrule{0.01em}{2pt}{2pt}
all                      & each, every, any, whole, entire,   total                                                                                               \\ \specialrule{0.01em}{2pt}{2pt}
group &
  \begin{tabular}[c]{@{}l@{}}batch, organize, categorize,   classify, arrange, separate, \\ label, tag, mark, pack, collect, assemble,  distribute,  \\ gather, merge, put together, index, concentrate, combine\end{tabular} \\ \specialrule{0.01em}{2pt}{2pt}
Sort                     & order, rank, sequence                                                                                                                  \\ 
\bottomrule
\end{tabular}%
}
\caption{Replacement rules for paraphrasing NL explanation}
\label{tab:table_replacement}
\end{table*}

\section{Synonym Substitution Rules for Paraphrasing}
\label{app:replacement}
To increase the NL explanation diversity in our training dataset, we paraphrase each machine-generated explanation by randomly replacing the NL explanation template words with substitute synonyms listed in Table~\ref{tab:table_replacement}. For example, the machine-generated explanation ``{\em return name}'' can be paraphrased to ``{\em find name}'' by replacing ``{\em return}'' with ``{\em find}''.

\section{Experiment Setup \& Hyperparameters}
We run our experiment on a server with Ubuntu 20.04, 2 NVIDIA Tesla T4 GPUs (16 GB), Intel Core i7-11700K GPU, and 64 GB memory. 

For the text-to-clause model, we follow the same architecture of SmBoP \citep{smbop}. Specifically, our model consists of 24 transformer layers, followed by another 8 RAT-SQL \citep{ratsql} layers. Each transformer has 1 feed-forward layer, 8 attention heads, and 256 dimensions. For each user-given NL question, it is encoded together with the database schema using GRAPPA \citep{grappa}. 

We finetuned the text-to-clause model and selected the best-performing model with the following hyperparameters: {\em optimizer} = {\em Adam}, {\em learning rate} = $1.8e-4$, {\em dropout rate} = $0.1$, {\em beam size} = $26$, {\em epoch} = 240,  {\em batch size} = 12.

\section{The Impact of Paraphrasing on Model Performance}
\label{text-to-clause_model}
To investigate the impact of paraphrasing on model performance, we trained and tested the text-to-clause models under 3 conditions: (1) the explanation is generated by {\tool} and not paraphrased, (2) the machine-generated explanation is paraphrased by the replacement rules in Table \ref{tab:table_replacement}, and (3) the machine-generated explanation is paraphrased by QuillBot. Then we evaluate the exact set matching match accuracy of generated clauses in Table~\ref{tab:paraphrase}.
Furthermore, we evaluate the end-to-end SQL generation accuracy in our user simulation experiment under 3 conditions, as shown in Table~\ref{tab:result3}. 
Overall, paraphrasing does not greatly impact the performance of text-to-clause SQL. 





\begin{table}[!h]
\centering
\resizebox{0.4\textwidth}{!}{%
\begin{tabular}{@{}lc@{}}
\toprule
  &\textbf{Acc$_{\scriptsize \rm  \textbf{set}}$}                     \\
  \midrule

No paraphrasing     & 0.922                                                  \\       
Paraphrasing with synonym substitution        & 0.915                                        \\
Paraphrasing with QuillBot   & 0.906                                               \\

                                       \bottomrule
\end{tabular}%
}
\caption{The exact set matching accuracy of the text-to-clause model when trained with three different datasets.} 
\label{tab:paraphrase}
\end{table}


\begin{table}[!h]
\centering
\resizebox{0.36\textwidth}{!}{%
\begin{tabular}{lcc} 
\toprule              
\specialrule{0em}{0pt}{1pt}
& Acc$_{set}$ & Acc$_{exec}$  \\ 
\specialrule{0em}{0pt}{1pt}
\midrule
\specialrule{0em}{0pt}{2.5pt}

\begin{tabular}[c]{@{}c@{}}SmBoP+STEPS$^{unpara}$\\
\end{tabular} & 0.981              & 0.973               \\

\specialrule{0em}{0pt}{1pt}
\begin{tabular}[c]{@{}c@{}}SmBoP+STEPS$^{substitute}$\\
\end{tabular}   & 0.975              & 0.973              \\ 

\specialrule{0em}{0pt}{1pt}

\begin{tabular}[c]{@{}c@{}}SmBoP+STEPS$^{quillbot}$\\
\end{tabular} & 0.975              & 0.971               \\

\specialrule{0em}{0pt}{1pt}

\bottomrule
\end{tabular}}
\caption{The end-to-end SQL generation accuracy of {\tool} when using the text-to-clause model trained on different datasets.}
\label{tab:result3}
\end{table}

\section{Direct Transformation Algorithm}
\label{appendix:Algorithm_direct_transformation}

As mentioned in Sec.~\ref{Sec:direct_transformation}, we define three types of {\em atomic edits}. 
While one can always design new transformation rules to support other simple edits, here we focus on these three basic edit types in order to demonstrate the benefits of direct transformation.  

Algorithm~\ref{algo:direct_transformation} describes the direct transformation algorithm. 
First, {\tool} performs chunking on the original explanation $e_o$ and the user-corrected explanation $e_n$ (Line \ref{line:1}-\ref{line:2}). We choose to split an explanation into phrases rather than individual words in order to recognize column names and table names that are represented as compound nouns in an explanation. Then, the chunks are aligned using the \citet{needleman} algorithm. If a chunk from the original explanation is aligned with a chunk in the new explanation and both of them can be mapped to a column name, a table name, or a literal value, then {\tool} replaces the corresponding name/value from the original clause with the new name/value (Line \ref{line:4}-\ref{line:5}). If a chunk from the new explanation is aligned with nothing, the chunk can be mapped to a column name, and the original clause is a \texttt{SELECT} clause, then {\tool} directly appends the corresponding column name to the clause after a comma (Line \ref{line:6}-\ref{line:9}). If a chunk from the old explanation is aligned with nothing and the chunk can be mapped to a column name, then {\tool} directly removes the corresponding column name from the clause (Line \ref{line:10}-\ref{line:12}).

\section{Impact of SQL Expertise}

\begin{figure}
    \centering
    \includegraphics[width=0.9\linewidth]{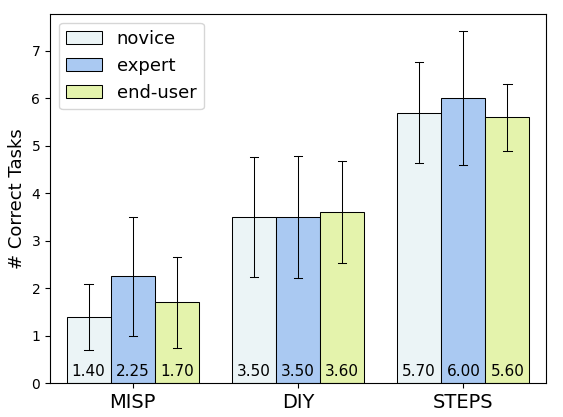}
    \caption{Tasks correctly completed by users with different levels of SQL expertise.}
    \label{fig:expertise}
\end{figure}

Figure~\ref{fig:expertise} shows a performance breakdown in our user study based on participants' SQL expertise.
We observe that expertise does not impact performance, with consistent performance on all tools by all groups.
The means for MISP do differ, but the distributions of scores overlap substantially.

\end{document}